\documentclass[10pt,letterpaper]{article}
\usepackage[top=0.85in,left=2.75in,footskip=0.75in]{geometry}

\usepackage{amsmath,amssymb}

\usepackage{changepage}

\usepackage[utf8x]{inputenc}

\usepackage{textcomp,marvosym}

\usepackage{cite}

\usepackage{nameref,hyperref}

\usepackage{microtype}
\DisableLigatures[f]{encoding = *, family = * }

\usepackage[table]{xcolor}

\usepackage{array}

\newcolumntype{+}{!{\vrule width 2pt}}

\newlength\savedwidth

\raggedright
\usepackage[aboveskip=1pt,labelfont=bf,labelsep=period,justification=raggedright,singlelinecheck=off]{caption}

\makeatletter
\renewcommand{\@biblabel}[1]{\quad#1.}
\makeatother

\usepackage{lastpage,fancyhdr,graphicx}
\usepackage{epstopdf}
\fancyheadoffset[L]{2.25in}
\fancyfootoffset[L]{2.25in}
\def\snr{\mathit{SNR}}
\def\raw#1{{\widetilde{#1}}}

\begin{document}
\vspace*{0.2in}

\begin{flushleft}
{\Large
  \textbf\newline{
    BCI-Walls: A robust methodology to predict if conscious EEG changes can be detected in the
    presence of artefacts
}}
\newline
\\
Bernd Porr\textsuperscript{*1\Yinyang},
Luc\'ia Mu\~noz Bohollo\textsuperscript{1\Yinyang}
\\
\bigskip
\textbf{1} Biomedical Engineering, University of Glasgow, Glasgow, Scotland, UK
\\
\bigskip

% Insert additional author notes using the symbols described below. Insert symbol callouts after author names as necessary.
% 
% Remove or comment out the author notes below if they aren't used.
%
% Primary Equal Contribution Note
\Yinyang These authors contributed equally to this work.

% Use the asterisk to denote corresponding authorship and provide email address in note below.
* bernd.porr@glasgow.ac.uk

\end{flushleft}

\section*{Abstract}
Brain computer interfaces (BCI) depend on reliable realtime detection
of conscious EEG changes for example to control a video game. However,
scalp recordings are contaminated with non-stationary noise, such as
facial muscle activity and eye movements. This interferes with the
detection process making it potentially unreliable or even
impossible. We have developed a new methodology which provides a hard and measurable criterion
if conscious EEG changes can be detected in the presence of
non-stationary noise by requiring the signal-to-noise ratio of a scalp
recording to be greater than the SNR-wall which in turn is based on the highest and
lowest noise variances of the recording. As an instructional example,
we have recorded signals from the central electrode Cz during
eight different activities causing non-stationary noise such as
playing a video game or reading out loud. The results show that facial
muscle activity and eye-movements have a strong impact on the
detectability of EEG and that minimising both eye-movement artefacts and muscle
noise is essential to be able to detect conscious EEG changes.

%\linenumbers

% Use "Eq" instead of "Equation" for equation citations.
\section*{Introduction}
Brain computer interfaces (BCI) have broadly the task to turn brain
activity (EEG) into actions, for example, to control a character in a
video game or a wheelchair. In the simplest way this is achieved by
testing if the EEG signal has reached a certain threshold
\cite{Pfurtscheller2006}. However, it is well known that
noise originating from muscle activity, eye-movements and other
movement artefacts can interfere with the detection of conscious EEG
changes \cite{Goncharova2003}. Various techniques have been
devised to minimise the effect of noise, in particular the independent
component analysis (ICA) for offline processing \cite{Delorme2007} but
for realtime closed-loop applications such as BCI one needs to resort to direct
causal filtering techniques using bandpass filters, the short-time
Fourier Transform, wavelet transform
\cite{Ahmadi2012,jirayucharoensak2013online,Jirayucharoensak2019} or
the derivative \cite{duvinage2012biosignals}. All
these approaches are inferior to the offline noise removal techniques
such as ICA and even after filtering the remaining noise will substantially
interfere with the detection process.

\begin{figure}[!hbt]
\begin{center}
\mbox{\includegraphics[width=\textwidth]{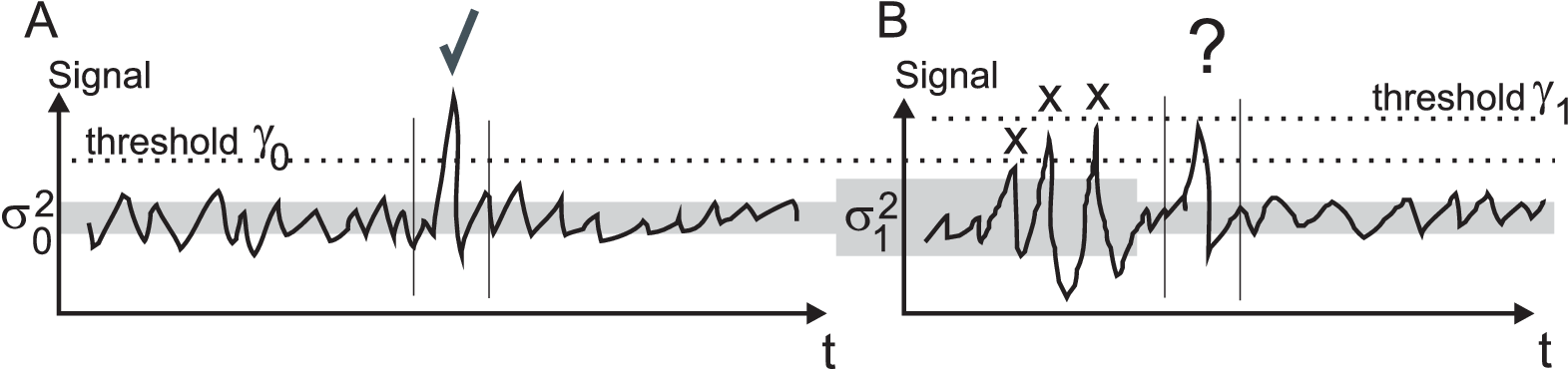}}
\end{center}
\caption{\textbf{Effects of stationary versus non-stationary noise on signal detection.}
  A) Signal detection with stationary noise, B) signal detection with non-stationary noise
  \label{stat_nonstat}}
\end{figure}

Given the limited effectiveness of noise reduction techniques for BCI
applications, electrode measurements will be substantially
contaminated with noise, in particular EMG, because of the overlap in
EEG and EMG frequencies, no matter what kind of pre-filtering has
been applied. Consequently, the detection process itself needs to cope
with the noise and minimise its impact. The standard solution to
maximise the signal and minimise the noise during detection is
\textsl{averaging} which can be achieved in both the time domain and
frequency domain:
\begin{itemize}
\item Time-domain: In the ``evoked potential'' (ep) paradigm the
  EEG is stimulus-locked and assumes that EMG
  noise is uncorrelated to the stimulus repetition and
  averages out.
\begin{equation}
\textrm{ep}[m] = \frac{1}{N} \sum_{n=0}^N d[m+n \cdot N]
\label{vep}
\end{equation}
where $N$ is the number of stimulus repetitions and their responses
registered as $d[m+n \cdot N]$. The more repeated stimuli $N$ are
presented the more the EMG noise is reduced. For example in a P300 speller
a subject looks at a flashing ``A'' and the EEG is then added
over and over again until a threshold has been reached. The
more repetitions of the letter ``A'' the better the signal-to-noise ratio (SNR)
but the longer the time to bring it over a threshold to decide the
user has looked at the flashing ``A''.
\item Frequency-domain: Here, the idea is that the subject can
  consciously reduce (or sometimes increase) the power of a narrow
  frequency band. To detect this change the signal is analysed in the
  frequency domain. If the band-power of a frequency band reaches a
  certain threshold then an action can be triggered for example moving
  a cursor. Again, averaging takes place because the Fourier Transform
  or a bank of bandpass filters accumulate the correlation between
  sine/cosine-waves ($e^{-j2\pi kn / N}$) and chunks of EEG $d[n]$:
\begin{equation}
X[k] = \sum_{n=0}^{N-1} d[n] \cdot e^{-j2\pi kn / N} \qquad k=0,1,2, \ldots, N-1
\label{DFT}
\end{equation}
where $N$ is the number of samples the averaging takes place,
$d[n]$ is the EEG and $X[k]$ its spectrum. If the
chunk of EEG is long then the frequency spectrum will deliver clear
peaks in the band of interest and thresholding becomes more and
more reliable.
\end{itemize}
No matter if the detection process is performed in the time- or
frequency-domain one needs to wait for $N$ samples until a decision
can be made so that a signal reaches a threshold.
Fig.~\ref{stat_nonstat}A shows such a case where a cartoon signal
is shown where a signal has to reach the threshold $\gamma$ which
then can be used to control for example a cursor in a BCI game.
The threshold $\gamma_0$ is chosen in a way that the noise with
noise variance $\sigma_0^2$ does not reach the threshold but the
desired conscious EEG signal does (indicated with the tick symbol).

In classical detection theory there is always a number of
samples $N$, if averaged over these, where reliable detection is
possible. However, this does not take into account
\textsl{non-stationary} noise which is the case when recording EEG
contaminated with EMG, eye-movements and movement artefacts
(Fig.~\ref{stat_nonstat}B). Here, the noise
variance changes, for example, from the small noise variance
$\sigma^2_0$ to a large noise variance $\sigma^2_1$ and then back to
the small noise variance $\sigma^2_0$. The three peaks indicated at
``X'' are noise peaks. The threshold $\gamma_1$ could be set higher
than $\sigma^2_0$ so that the 2nd and 3rd noise peaks are not
detected. However, now we are encountering a problem at ``?'' which
could be either noise or signal. It is most likely an actual signal as
the noise variance has just dropped again to a lower level but this
can only be known because noise levels have been asserted
\textsl{a priori}. One could argue again that averaging over
more samples $N$ will eventually average out the noise but because of
the changing noise variance peaks as the ``?'' become
\textsl{ambiguous} and can either be signal or noise. Analytically
this means that with non-stationary noise the number of samples $N$
required to detect robustly a signal can reach \textsl{infinity} ($N
\to \inf$) and thus detection is not possible at all. This condition
is called \textsl{SNR-wall} \cite{Tandra2008}.

Facial EMG and
eye-movements are certainly non-stationary noise sources -- in
particular in everyday situations when subjects walk, play a video
game or simply talk. This means that any activity a subject does
will define a hard SNR-wall because the EMG and eye-movements are
non-stationary. If the SNR is below the SNR-wall it is not possible to
detect the EEG at all and thus it is not possible at all to design a
BCI system for that kind of activity.

In this paper we introduce the concept of \textsl{SNR-walls} to EEG
measurements to have an objective way to determine if an experiment can
detect EEG in principle in the presence of non-stationary noise. As
an instructional example of how to generally calculate SNR-walls, we
have recorded the EEG from subjects during a
range of different tasks. We will then calculate the SNR-walls for the
different activities and conduct a statistical analysis to gauge if it's possible to construct a BCI
system given the non-stationary noise created by these tasks.

\section*{Methods}
\subsection*{SNR-walls}
In this section, we briefly explain the relevant analytics of SNR-walls
and how to calculate them practically. Tandra et al \cite{Tandra2008}
provide the analytical derivation of the methodology and while they
applied it to the telecommunication domain we apply it to BCI.

Let us consider a signal measured with an electrode $\raw{d}[n]$ placed
on the head of a subject:
\begin{equation}
  \raw{d}[n] = \underbrace{\raw{a}[n] + \raw{b}[n]}_{\raw{r}[n]} + \raw{c}[n]
\end{equation}
where $\raw{c}[n]$ is the consciously controlled part of the
EEG. $\raw{b}[n]$ is the background EEG activity which the subject
cannot control. $\raw{a}[n]$ are all artefacts such as muscle activity
added to the measurement.
Important for this paper is that the noise sources
$\raw{a}[n]$ and $\raw{b}[n]$ are both non-stationary: for example
when smiling the power of $\raw{a}[n]$ will be larger and when
relaxing the power of $\raw{a}[n]$ will be less and non-stationary
muscle noise is even generated during pure mental tasks \cite{Whitham2008}. The only stationary noise in
these recordings comes from the measurement equipment. 
Together $\raw{b}[n]$ and $\raw{a}[n]$ form the noise
$\raw{r}[n]$ of the signal measured. Generally, the raw
EEG signal from the electrode requires filtering, which includes at
least DC and powerline interference removal but usually also bandpass
filtering if only a certain EEG band is of interest:
\begin{equation}
  d[n] = ( \raw{r}[n] + \raw{c}[n] ) * f[n]
\end{equation}
where we subsume all filtering in $f[n]$ and from now on signals without the $\;\raw{\,}\;$
are the ones after filtering.

The decision problem can be considered as a binary
hypothesis testing problem which can be written as:
\begin{eqnarray}
H_0:d[n] & = & r[n] + 0\\
H_1:d[n] & = & r[n] + c[n]
\end{eqnarray}
The hypothesis $H_0$ represents the situation where the signal at
the electrode contains just noise and $H_1$ where
consciously generated EEG is present.

Central to the BCI system is that it is able to detect the conscious EEG $c[n]$
so that an action can be generated, for example, steering
a wheelchair. Thus, we need a detector which takes the signal $d[n]$ and
decides with a threshold $\gamma$ if it has just been noise or consciously 
generated EEG on top of the noise.

If one knows nothing about the signal except that it will increase or
decrease one can just detect the average power over $N$ samples
and gain a test statistic:
\begin{equation}
  T(X) = \frac{1}{N} \sum_{n=1}^{N} x[n]^2 \label{power}
\end{equation}
which creates a random
variable $T$ which in turn is then compared
against a threshold $\gamma$ which decides if the signal has
contained the consciously controlled EEG component $c[n]$ or not.

Since the SNR-wall analytics is quite complex we follow the same gradual approach
Tandra et al used in their paper \cite{Tandra2008} and present the analytics in two steps:
first, we assume that the noise is stationary which leads to standard
detection theory and then in the 2nd step, we
extend these equations by taking into account non-stationary noise which leads to our new BCI-walls methodology.

Let us first assume that our total noise/artefacts $r[n]=a[n]+b[n]$ have a 
single nominal variance $\sigma_r^2$ which means that we have stationary noise.
We calculate our detection probabilities $P(D)$ as:
\begin{eqnarray}
  P(D)|H_0 & \sim &
  \mathcal{N}\left(0 +\sigma_r^2,\frac{1}{N}2(0 +\sigma_r^2)^2 \right) \\
  P(D)|H_1 & \sim &
  \mathcal{N}\left(T(c) +\sigma_r^2,\frac{1}{N}2(T(c)+\sigma_r^2)^2 \right)
\end{eqnarray}
where $c$ is the consciously controlled EEG power.

The SNR of the signal $d(n)$ can be expressed as:
\begin{equation}
  \snr = \frac{T(c)}{\sigma_r^2} \label{snrdef}
\end{equation}
where $T(c)$ is the average EEG power of the consciously generated
EEG component $c[n]$ and $\sigma_r^2$ is our nominal noise
variance (i.e. noise power).

The probability of a false alarm $P_{FA}$ can be written as:
\begin{equation}
P_{FA} = \mathcal{Q}\left( \frac{\gamma - \sigma_r^2}
                        {\sqrt{\frac{2}{N}}\sigma_r^2} \right) \label{pfa}
\end{equation}
where $N$ is the number of samples, $\sigma_r^2$ the noise power,
and $\gamma$ the detection threshold.

In a similar way, the probability for detection $P_D$ is given by:
\begin{equation}
P_D = \mathcal{Q}\left( \frac{\gamma - (T(c)+\sigma_r^2)}
                    {\sqrt{\frac{2}{N}} (T(c)+\sigma_r^2)}
        \right) \label{pd}
\end{equation}
By eliminating the detection threshold
$\gamma$ in Eq.~\ref{pfa} and Eq.~\ref{pd} and with the
help of Eq.~\ref{power} and Eq.~\ref{snrdef} we are able to obtain an 
equation for the number of samples $N$ required to detect robustly the 
conscious EEG component $c$:
\begin{equation}
N = \frac{2[\mathcal{Q}^{-1}(P_{FA}) -
  \mathcal{Q}^{-1}(P_D)(1+\snr)]^2}{{\snr}^2}
\end{equation}
which means that at a constant noise power $\sigma_r$ one just needs to
average over $N$ timesteps (Eq.~\ref{power}) to obtain a
robust detection and with sufficiently large $N$ it is always possible
to perform a robust detection.

However, brain recordings are contaminated by
non-stationary noise and we need to extend the above analytics
to take into account the random changes of the noise variance.
To capture this noise uncertainty we define the parameter $\rho$ which is
the ratio between the largest noise variance
$\sigma_{r,\textrm{max}}^2$ and the smallest one
$\sigma_{r,\textrm{min}}^2$:
\begin{equation}
  \rho = \sqrt{\frac{\sigma_{r,\textrm{max}}}{\sigma_{r,\textrm{min}}}}
  \label{rho}
\end{equation}

Our nominal noise variance $\sigma^2_r$ is then defined as:
\begin{eqnarray}
\sigma_{r,\textrm{min}}^2 & = & \frac{1}{\rho} \sigma^2_r \label{minsigma} \\
\sigma_{r,\textrm{max}}^2 & = & \rho \sigma^2_r \label{maxsigma}
\end{eqnarray}

Having defined the noise uncertainty we can express our false alarm and 
detection probabilities as a function of $\rho$:
\begin{eqnarray}
  P_{FA} & = & \mathcal{Q}\left( \frac{\gamma - \rho\sigma_r^2}{\sqrt{\frac{2}{N}}\rho\sigma_r^2} \right)
  \label{pfa_rho} \\
P_D & = & \mathcal{Q}\left( \frac{\gamma - \left( T(c)+\frac{1}{\rho}\sigma_r^2 \right)}
                    {\sqrt{\frac{2}{N}} \left( T(c)+\frac{1}{\rho}\sigma_r^2 \right)}
                    \right)
   \label{pd_rho}
\end{eqnarray}

In a low SNR environment, $\snr+1 \approx 1$
can be assumed and Eqs.~\ref{pfa_rho}, \ref{pd_rho} and \ref{power} combined yield \cite{Tandra2008}:
\begin{equation}
  N = \frac{ 2 \left[\mathcal{Q}^{-1}(P_{FA}) - \mathcal{Q}^{-1}(P_D)\right]^2}
           { \left[\snr-(\rho-\frac{1}{\rho})\right]^2}
  \label{noisewalltime}
\end{equation}
where $N$ is again the number of samples that are needed to achieve
the target probability of a permitted false alarm and probability of
detection. However, now we note that when the SNR decreases the required
integration/averaging steps $N$ become infinite at
$\snr = (\rho-\frac{1}{\rho})$. This critical SNR is called the \textsl{SNR-wall}. 
Since only the denominator of Eq~\ref{noisewalltime} counts, the SNR wall can
be calculated simply as:
\begin{eqnarray}
\snr_\textrm{wall} = \rho - \frac{1}{\rho} \label{snrwall}
\end{eqnarray}
The SNR-wall represents a fundamental limitation of detection
ability, which means that relevant uncertainties cannot be
countered by a longer averaging time. In order to determine if
EEG can be detected or not we need to determine both the noise
uncertainty $\rho$ and the SNR of our measured brain signal.

Having now derived the underlying analytics we can devise a step-by-step guide on 
how to determine if EEG changes $c[n]$ can be detected at all:
\begin{enumerate}
\item \textbf{Calculate the SNR-wall}
  by using the minimum noise variance
  $\sigma_{r,\textrm{min}}^2$ and the maximum noise variance
  $\sigma_{r,\textrm{max}}^2$ of the brain recording (Eq.~\ref{snrwall}).
\item \textbf{Calculate the SNR} as the ratio of pure consciously controlled EEG power and noise power
  (Eq.~\ref{snrdef}).
\item \textbf{Compare the SNR with the SNR-wall:}
   \begin{equation}
    \textrm{\mbox{conscious EEG changes are detectable}} = 
    \begin{cases}
      \textrm{yes} &  \snr > \snr_\textrm{wall} \\
      \textrm{no} & \text{otherwise}
    \end{cases}
  \end{equation}
\end{enumerate}
In the following sections we call the entire three-step process of determining if
conscious EEG can be detected ``BCI-Wall''.

Having described the analytical derivations of the BCI-wall in detail
we now show how to apply it in practise. This also acts
as an instructional example of how to determine the BCI walls for other experiments.

\subsection*{Data acquisition}
As shown above, in order to determine if conscious EEG changes can be
detected one needs to calculate the SNR-wall and the SNR. The SNR-wall
depends on the \textsl{ratio} between minimum noise variance and
maximum noise variance (Eq.~\ref{snrwall}) where different activities
will cause different noise variances. For example lying down with eyes
closed will have quite similar maximum and minimum noise variances
resulting in a low SNR-wall. However, while reading out loud
facial muscles will create short bursts of strong muscle noise as
well as eye movement artefacts and the ratio between smallest and
highest noise variance will be large which in turn will result in
a high SNR-wall. Consequently, we have devised different
tasks ranging from low noise variance ratios while lying down to
extreme noise variance ratios during jaw clench. The resulting
SNR-wall is then compared with the SNR. This is calculated by using the
P300 evoked potential as a measure of the consciously
generated EEG power and dividing it by the nominal noise power.
We are now describing the experimental procedure.

Data was obtained from 20 healthy participants (9 males, 11
females). Prior to the experiment, participants were given an
information sheet and were asked to give signed consent by signing two
consent forms, one for the researchers and another for them to
keep. Ethical approval was given by the ethics committee at the Institute of
Neuroscience and Psychology, School of Psychology at the University of
Glasgow, with reference 300210055. The data was acquired using an
Attys data acquisition device (\url{www.attys.tech}), made up of 2
channels at 24-bit resolution, and its data acquisition programmes
`attys-ep' and `attys-scope'. EEG recordings were single channel
between an Ag/AgCl electrode at Cz \& A2 and GND at A1. The 2nd channel was not used.

Participants 2 and 6 had to be excluded from the study because of faulty electrodes.
About 180 sets of data are analysed in this study. These are stored in an open-access 
database \cite{Bohollo2022}, all ethics files can also be found on the database.

Data was recorded during 10 activities and P300 where pink horizontal
stripes were produced randomly after 7 to 13 seconds while otherwise a
chequerboard was inverted every second.  The P300 was recorded for a
duration of 5 minutes while the participants sat on a chair directly
opposite a screen. The tasks generating various levels of artefacts
were jaw clenching, reading, colouring, attempting a word search,
trying a Sudoku, playing `Subway Surfers' game on the phone, lying
with eyes closed and repeat with eyes opened. Subjects sat for all
tasks except the lying down ones. Each of these ran for a duration of
2 minutes. The `attys-ep' programme was used to record the P300, while
the rest were recorded using `attys-scope'.

\subsection*{EEG pre-processing}
The raw EEG data $\raw{d}[n]$ undergoes causal filtering prior to detection.
Here, we have subsumed all pre-processing steps in the
filter function $f[n]$. We are presenting
five different filtering setups:
\begin{enumerate}
\item Wideband detector with minimal filtering: $f[n]$ =
  $0.1~\mathrm{Hz}$ 4th order highpass and 50~Hz powerline bandstop which
  preserves the entire EMG spectrum and also low frequency
  fluctuations such as eye-movements. This pre-processing is hardly used in practise but acts here as
  a worst-case scenario as the whole EMG spectrum is allowed to
  interfere with the detection.
\item Low frequency bandpass filtering between a $0.1-3~\mathrm{Hz}$
  4th order Butterworth filter and 50~Hz powerline
  bandstop. This frequency range is in the EEG delta frequency range
  and is in particular suitable to investigate the effect of low
  frequency artefacts on the ability to detect EEG such as eye-blink,
  eye-movements and movement artefacts from the cables.
\item Wideband bandpass: $f[n]$ = 4th order Butterworth bandpass
  $8-18~\mathrm{Hz}$, DC-removal, 50~Hz powerline bandstop with a moderate
  rejection of higher EMG frequencies used for motor imagination
  \cite{PFURTSCHELLER2006153}.
\item Narrow bandpass: $f[n]$ = 4th order Butterworth bandpass
  $8-12~\mathrm{Hz}$, DC-removal, 50~Hz powerline bandstop detecting alpha power
  around 10~Hz and rejects the higher frequency EMG power
  \cite{Wolpaw1991}.
\item Difference operator: $f[n] = f[n] - f[n-1]$ and 50~Hz powerline bandstop which has been used
  for example in \cite{duvinage2012biosignals}.
\end{enumerate}

These five post-processing scenarios are applied separately to the data of all subjects and tasks, 
except for those with obvious broken electrode signals or strong artefacts.

\subsection*{BCI-wall calculation}

As outlined above, the BCI-wall calculations require three steps: 1) SNR-wall calculation,
2) SNR calculation and 3) comparing SNR and SNR-wall to determine if conscious
EEG detection is possible. We are now describing how this can be done practically
and is also an instructional example for other datasets. The corresponding Python code
is available on GitHub\cite{bernd_porr_2023_7852162}.

\begin{figure}[!hbt]
\begin{center}
\mbox{\includegraphics[width=\textwidth]{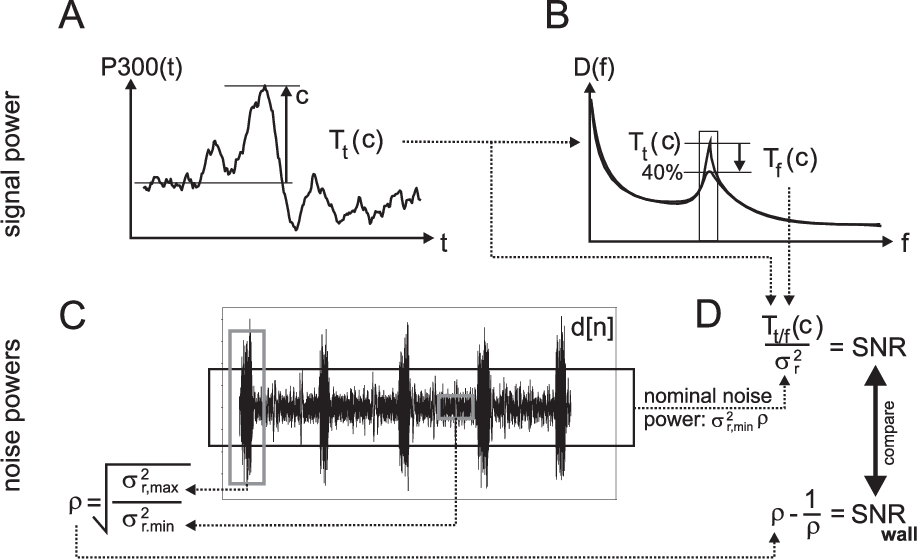}}
\end{center}
\caption{\textbf{Flow diagram illustrating the calculation of the SNR and SNR-wall.} A) Time domain
  signal for the SNR calculation takes the power $T_t(c)$ of the P300 peak as the consciously
  generated EEG change. B) Frequency domain signal power for the SNR calculation
  takes the power $T_t(c)$ and assumes a conscious 40\% reduction of the signal power
  in a narrow band around a peak. C) For the overall SNR the nominal variance $\sigma^2_r$ is calculated
  over the whole EEG recording. To calculate the SNR-wall the maximum $\sigma^2_{r,\textrm{max}}$ 
  variance and minimum $\sigma^2_{r,\textrm{min}}$ variance is detected with a sliding window 
  sample by sample. D) If the SNR is greater than the SNR-wall then a conscious change in the EEG 
  can be detected. If the SNR is less than the SNR-wall then it is impossible to detect a conscious 
  change.
  \label{calcflow}}
\end{figure}

\subsubsection*{Step~1: SNR-Wall}
In order to calculate the SNR-wall we need to calculate $\rho$ (Eq.~\ref{rho}) which is
the \textsl{ratio}
between the maximum $\sigma_{r,\textrm{max}}^2$ and minimum $\sigma_{r,\textrm{min}}^2$ noise power.
To find these two values we use a sliding window of length $\tau = 2~\mathrm{sec}$ and
calculate the noise power for different sample positions:
\begin{equation}
  \sigma_{\textrm{chunk}}[n] = T(d[n \ldots n+\tau])
\end{equation}
and obtain its maximum and minimum over the entire recording of $N$ samples:
(see Fig.~\ref{calcflow}):
\begin{eqnarray}
\sigma_{r,\textrm{min}}^2 & = & \min_{n=0\ldots N-1}{\sigma_{\textrm{chunk}}[n]} \label{minvar} \\
\sigma_{r,\textrm{max}}^2 & = & \max_{n=0\ldots N-1}{\sigma_{\textrm{chunk}}[n]} \label{maxvar}
\end{eqnarray}
With Eqs.~\ref{rho} and \ref{snrwall} one can then calculate the
SNR-wall which is usually expressed in dB (to make it comparable to the
SNR which is calculated in the next section):
\begin{eqnarray}
\snr_\textrm{wall} = 10 \log_{10} \left( \rho - \frac{1}{\rho} \right) \label{snrwalldB}
\end{eqnarray}

\subsubsection*{Step~2: SNR}
The SNR is calculated as a ratio between signal power $T(c)$ and the nominal noise power
$\sigma_r^2$:
\begin{equation}
  \snr = 10 \log_{10} \left( \frac{T(c)}{\sigma_r^2} \right ) =
         10 \log_{10} \left( \frac{T(c)}{\rho\sigma_{r,\textrm{min}}^2} \right )
  \label{snragain}
\end{equation}
where we can re-use the SNR-wall parameters $\rho$ and $\sigma_{r,\textrm{min}}^2$ from
the previous section to calculate the nominal noise power $\sigma_r^2$.
Measuring the pure conscious changes of EEG $c$ and its power $T(c)$ is only indirectly 
possible as it is not ethical to paralyse subjects. The P300 evoked potential offers
an individual estimate of the consciously generated signal power $T(c)$ calculated 
from the peak P300 voltage. This will be used for estimating the power of the signal 
for both a time domain task (i.e. P300) and frequency domain task (i.e. motor 
imagination):
\begin{itemize}
\item \textbf{Time-domain power of the signal:} The P300 peak can be used straight away
for a time domain calculation of its power (Fig.~\ref{calcflow}A):
\begin{equation}
  T_t(c) = c_\textrm{max}^2 \label{signalpowertime}
\end{equation}
where the $T_t(c)$ is the power of the pure EEG in the time domain.
\item \textbf{Frequency-domain power of the signal:}
BCI systems using the frequency domain change the power of a narrow
frequency band (Fig.~\ref{calcflow}B), for example by motor imagination. Here, we assume that:
\begin{itemize}
\item The EEG power generated in a narrow frequency band is comparable
  to the power generated in the time domain $T_t(c)$. This is
  indicated between the panels Fig.~\ref{calcflow}A and B by noting
  that the peak powers are identical.
\item However, because motor imagination \textsl{reduces} power
  consciously the actual conscious power $T_f(c)$ is only as
  strong as the \textsl{reduction} of power. Here, we assume a
  40\% reduction of power.
\end{itemize}
Given that only the reduction is the conscious power change $c$ and anything
else will be absorbed in the noise term $r$ we
can calculate the pure signal power (in contrast to the noise) as:
\begin{equation}
  T_f(c) = T_t(c) - T_t(c) \cdot 40\%
\end{equation}
\end{itemize}

After having calculated the power of the signal $T_f(c)$ we now need
to calculate the power of the noise. Given that the small P300 evoked
potentials $c$ are buried in the EEG $d$, one can assume that the noise
variance containing a consciously generated signal and one without are
basically identical \cite{Hu2010}: $\sigma_r^2 \approx
\sigma_d^2$. Thus we take the variance of the EEG epoch $d[n]$
(Fig.~\ref{calcflow}D) as
\begin{equation}
  \sigma_r^2 \approx T(d)
\end{equation}
where the power of the EEG epoch $T(d)$ is calculated with Eq.~\ref{power}
which is the average power or variance.

\subsubsection*{Step~3: Comparing SNR and SNR-Wall:
  determining if conscious control can be detected}
If the SNR of the EEG (Eq.~\ref{snrdef}) is above the SNR-wall
(Eq.~\ref{snrwall}) then detection of the conscious EEG change $c$ is
possible. Otherwise not.

For every task, for example ``Sudoku'', there will be individual pairs
of SNR and SNR-wall values from every subject. Because the SNR and SNR-wall
values are calculated over all subjects they are random variables. A
t-test is used to determine if the SNR for all subjects is
significantly above the SNR-wall for each task and each of the
four post-processing scenarios.

\section*{Results}
We are going to describe the results in three steps:
\begin{enumerate}
\item\textbf{Investigating the raw data:}
  The different datasets (jaw clench, lying eyes closed/open, word search,
  Sudoku, phone app, reading, colouring) used in this
  study exhibit different signal and noise characteristics in their time- and frequency
  domain. We will use subject \#20 as a \textsl{representative example} to point
  out the differences between the datasets.
\item\label{res2}\textbf{Walk-through of the calculation of the SNR and SNR-wall
  of one subject for both jaw clench and reading:}
  As instructional examples we calculate step by
  step the SNRs and SNR-walls of \textsl{one} subject for
  \textsl{jaw-clench} and \textsl{reading} at different EEG frequency
  ranges such as: full range, only low frequencies ($0.1-3~\mathrm{Hz}$),
  wide bandpass ($8-18~\mathrm{Hz}$),
  narrow bandpass ($8-12~\mathrm{Hz}$) and derivative.
\item \textbf{Statistical analysis with t-test:}
  Using the same approach as in step~\ref{res2} but for \textsl{all} subjects enables us to do
  a statistical analysis with a t-test which determines if the SNR of a task is \textsl{significantly}
  larger than the SNR-wall and thus conscious EEG changes can significantly be
  detected.
\end{enumerate}

\begin{figure}[!hbt]
\begin{center}
\mbox{\includegraphics[width=\textwidth]{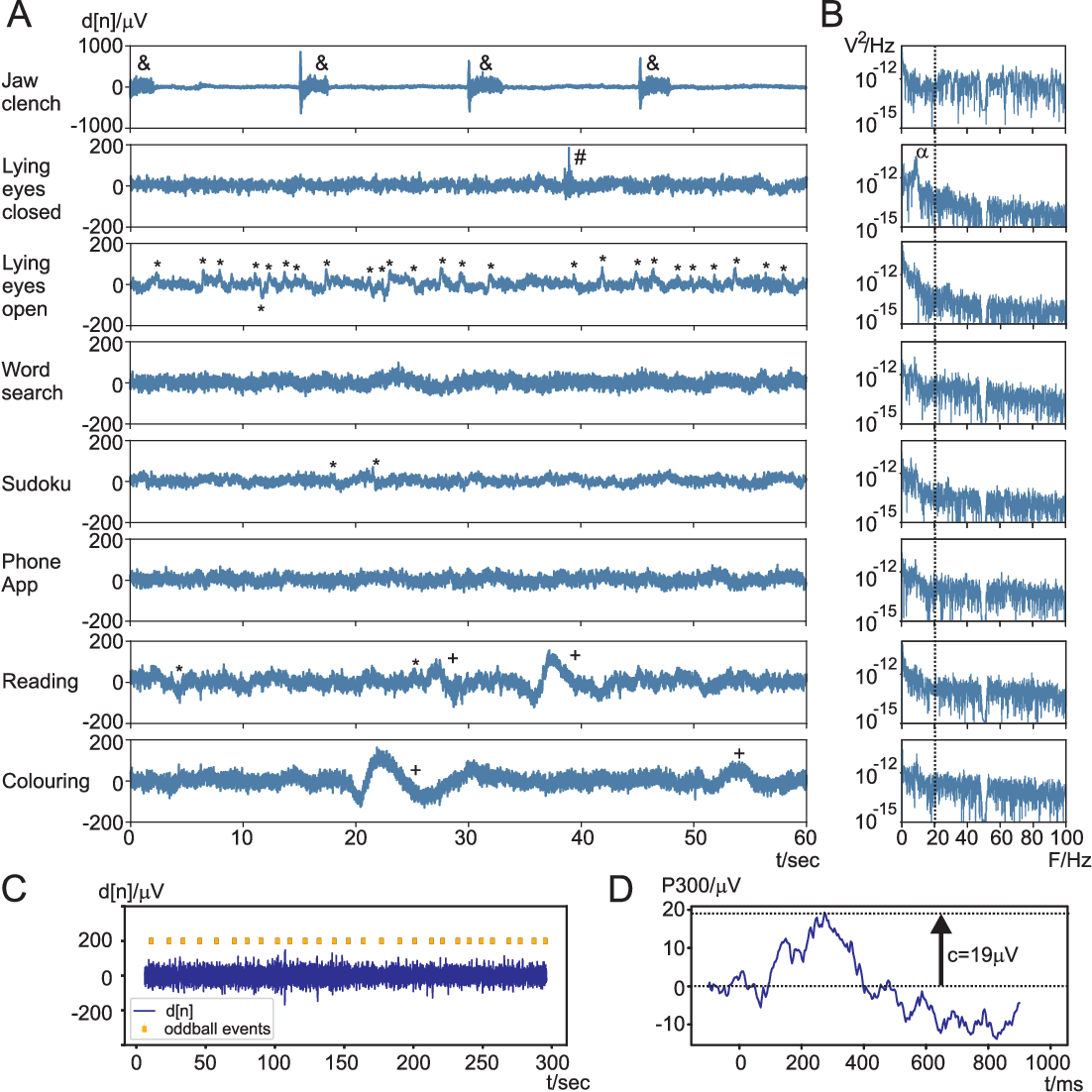}}
\end{center}
\caption{\textbf{Complete dataset of a representative recording
    (subject 20).}  A) Time domain of the different recordings ($d[n]$ in $\mu
  V$ against $s$). B) Power spectra of $d[n]$ of the corresponding recordings ($V^2/\mathrm{Hz}$).
  C) Raw P300 recording ($\mu V$ against $s$). D) Averaged P300 response ($\mu V$ against $ms$).
  \label{sub20}}
\end{figure}

\subsection*{Investigating the raw data}
Fig.~\ref{sub20} shows the measurement results taken from subject~20
which are representative of the artefacts encountered during the
different experimental tasks which are listed on the left-hand side:
jaw clench, lying eyes closed/open, word search, Sudoku, phone app,
reading, colouring. Fig.~\ref{sub20}A shows the time domain plots for
60 seconds. The signals plotted have only the DC and 50~Hz main
interference removed and otherwise represent the full range of raw
signals. Panel Fig.~\ref{sub20}B shows the power spectra in
logarithmic units of the corresponding time domain plot to the left in
Fig.~\ref{sub20}A. The dotted line at $20~\mathrm{Hz}$ marks the point
from where muscle noise contribution will decrease towards lower
frequencies \cite{Whitham2007} or in other words: we will expect
muscle noise for frequencies above $20~\mathrm{Hz}$ but also to a
lower degree down to $10~\mathrm{Hz}$. We are now going through
the different tasks step by step. We start with ``jaw clenching'' followed
by ``lying down, eyes closed'' being the tasks with the largest
difference in noise amplitudes where jaw clenching has about
50 times more noise than lying down. Note that, for the noise
wall calculation, it is the \textsl{ratio} between the highest
and lowest noise variances that are used (Eq.~\ref{snrwalldB}),
not the absolute amplitudes. The ratio is again highest for the
jaw clench:

\begin{itemize}
\item\textbf{Jaw clenching:} For this task the subject had to clench
  their jaw every 15 seconds when indicated by the researcher and these have been marked with an ``\&''
  in Fig.~\ref{sub20}A.
  Peak amplitudes at the onset of the jaw clench reach nearly $1~\mathrm{mV}$ while
  in the pauses the amplitude is just around $20~\mu\mathrm{V}$. Its corresponding spectrum in
  Fig.~\ref{sub20}B is almost flat over the whole frequency range.
  Compared to the relaxed state of
  lying down the spectral power during jaw clenching is approx 100
  times larger for frequencies above $20~\mathrm{Hz}$. This
  experiment is useful to gain an insight into which spectral components
  are affected by muscle noise and confirms that in particular
  frequencies above $20~\mathrm{Hz}$ are affected but this is of course
  no hard cutoff but the influence of muscle noise
  towards lower frequencies decreases gradually.
\item\textbf{Lying down, eyes closed:} Here, the subject was
  asked to relax as much as possible and since the eyes were closed
  there are no eyeblink artefacts. The amplitude is at about $20~\mu\mathrm{V}$ with
  small fluctuations. However, muscle activity can
  still occur as marked with ``\#'' where the subject had to swallow
  and the muscle activity can clearly be seen. In the frequency
  spectrum an alpha peak at $10~\mathrm{Hz}$ emerges which is expected
  when closing the eyes.
\item\textbf{Lying down, eyes open:} With the eyes open the alpha peak vanished and the subject
  was looking at different locations at the ceiling and in the lab
  which shows up as a mix of eye-blink and eye-movement artefacts
  marked with ``*''. Generally subjects tended to look around
  frequently as it is a slightly unusual situation.
\item\textbf{Word search:} Word search engages the eyes but only at
  a narrow viewing angle looking at a sheet of paper which causes only small
  eye saccades. However, the spectrum between $20-50~\mathrm{Hz}$
  is elevated indicating muscle activity,
  possibly because of tense muscles during the task.
\item\textbf{Sudoku} appears to have similar muscle activity as
  lying down with eyes closed, with little facial muscle tone. However,
  since the eyes were open and the subject had to scan the Sudoku
  grid a few sudden jumps in the EEG potential can be identified
  which were saccades indicated with a ``*''.
\item\textbf{Phone app game:} The phone app has a similar behaviour
  as ``word search'' in that the power spectrum is elevated
  $20-50~\mathrm{Hz}$ suggesting again high facial muscle activity.
\item\textbf{Reading aloud} creates a mix of artefacts as various
  facial muscles are needed for spoken word and the eyes need to scan
  the text which cause eye artefacts as well.  The frequency spectrum
  above $20~\mathrm{Hz}$ in Fig.~\ref{sub20}B is elevated and stays
  elevated up to $100~\mathrm{Hz}$. In addition lower frequency
  components from eye-blinks and saccades ``*'' appear towards
  $0.1\ldots 3~\mathrm{Hz}$. Very low frequency components indicated
  at ``+'' were generated because the subject was turning the page and
  then touching their cheek while reading.
\item\textbf{Colouring:} Here, the subject had to take
  different pens and colour a line drawing.
  This task is an activity where the colouring makes the whole
  body move in the rhythm of the strokes made with the pen.
  Interestingly these pen-strokes were in the region of 
  $10~\mathrm{Hz}$ and created a weak fake alpha peak in the spectrum.
  In addition the activity also created larger movement artefacts
  at ``+'' where the subject switched pens.
\end{itemize}

The final two panels in Fig.~\ref{sub20}C/D show the recording
of the P300 experiment. This experiment determines the \textsl{signal
  power} for the SNR calculation (see Eq.~\ref{signalpowertime}).
Recall that it is not ethical to paralyse a healthy subject
to determine their pure EEG signal power and for that reason
we use the P300 oddball reaction of the brain to have a conscious
voltage change to a surprising stimulus. The subject is presented
with the oddball event at irregular intervals as indicated in
Fig.~\ref{sub20}C as ``oddball events''. The actual change in EEG
is small and buried in the original electrode signal $d[n]$. After event triggered
averaging we arrive at Fig.~\ref{sub20}D where the oddball event
happened at $t=0~\mathrm{ms}$ and the EEG response then reached
its peak of $c_\textrm{max} = 19\mu\mathrm{V}$ after $300~\mathrm{ms}$.

\begin{figure}[!hbt]
\begin{center}
\mbox{\includegraphics[width=\textwidth]{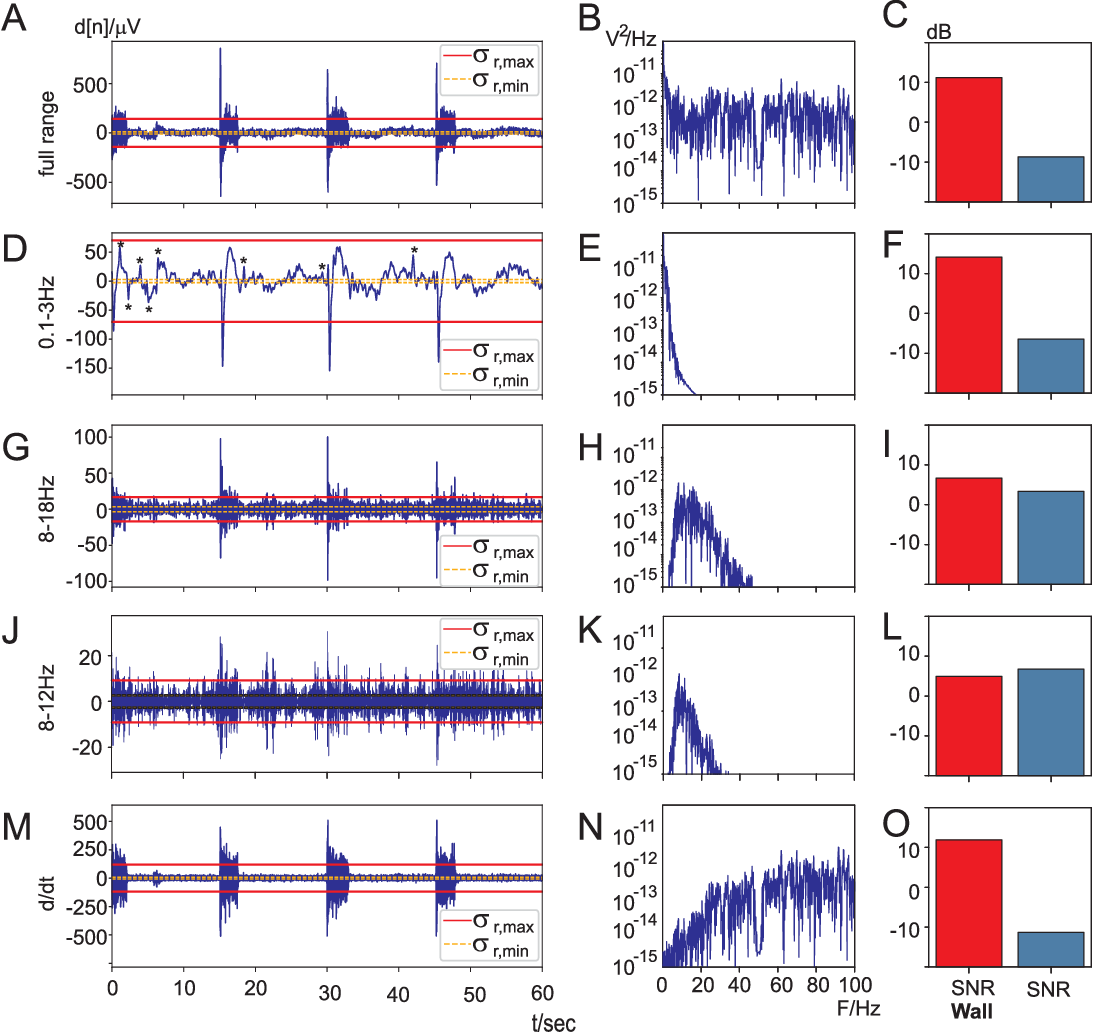}}
\end{center}
\caption{\textbf{SNR and SNR-wall calculations for the ``jaw clench'' task.}
  Columns A-M:
  electrode signals in the time domain with lines for the maximum variance $\sigma_{r,\textrm{max}}^2$
  and minimum variance $\sigma_{r,\textrm{min}}^2$ while using a sliding window of $2~\textrm{sec}$.
  Columns B-N: frequency power spectra
  of the corresponding time domain signals of columns A-M. Columns C-O:
  SNR-wall and SNR calculations of the corresponding time domain
  signals of columns A-M.
  \label{jaw}}
\end{figure}

\subsection*{Walk-through of the calculation of the SNR and SNR-wall
  of one subject for both jaw clench and reading}
We now present a walk-through of how to calculate the SNR and the
SNR-wall for the ``jaw clench'' task. This serves as an instructional
example for the statistical analysis for all subjects further
below. Fig.~\ref{jaw} shows the time- and frequency domain signals
during jaw clenching in panels A-M and B-N, respectively. 
Panels A and B are identical to the 1st row in Fig.~\ref{sub20}. The
column C-O shows the calculated SNR-wall and SNR values based on the
time domain data of the corresponding panels A-M. Let us first focus on
the first row Fig.~\ref{jaw}A-C and show step by step how to calculate
the SNR-wall and SNR for it. We start with the SNR-wall calculation
(Eqs.~\ref{rho}, \ref{snrwall} \& Fig.~\ref{calcflow}C,D) for which we
need the minimum variance (here: $\sigma_{r,\textrm{min}}^2 = 1.15
10^{-10} \mathrm{V}$) and maximum variance (here:
$\sigma_{r,\textrm{max}}^2 = 2.00 10^{-8} \mathrm{V}$).  This results
in $\rho = 13.21$ and an SNR-wall of $\snr_\mathrm{wall} =
11.18~\mathrm{dB}$ which is shown in Fig.~\ref{jaw}C. This means that
the SNR of the electrode signal needs to be larger than
$\snr_\mathrm{wall} = 11.18~\mathrm{dB}$ to be able to detect
conscious changes in the EEG. Consequently as a next step we are going
to calculate the SNR of the electrode signal.  The overall SNR of the
full range electrode signal is the ratio of the amplitude of the
conscious EEG change $c_\textrm{max}$ against the nominal noise power
(see Eq.~\ref{snragain}).  The conscious EEG change was already
determined in Fig.~\ref{sub20} as $c_\textrm{max} = 19~\mu\mathrm{V}$
and its power $T_t(c) = c_\textrm{max}^2 = 3.70
10^{-10}~\mathrm{V}^2$.  The nominal noise power $\sigma^2_r$ can
either be calculated with $\sigma^2_r = \rho
\sigma_{r,\textrm{min}}^2$ or directly from the electrode signal $d[n]$
(Eq.~\ref{snragain}) and results here in: $\sigma^2_r = 2.75
10^{-09}~\mathrm{V}^2$. This leads to an SNR of $-8.71~\mathrm{dB}$
which is shown in Fig.~\ref{jaw}C. Since the SNR is substantially
below the SNR-wall, detection of conscious changes of EEG are
\textsl{not} possible in this case.  This comes as no surprise as the
SNR-wall is determined by the \textsl{ratio} between the minimum noise
variance $\sigma_{r,\textrm{min}}^2$ and the maximum one
$\sigma_{r,\textrm{max}}^2$. The wider
the distance between these two lines in Fig~\ref{jaw}A the higher the SNR-wall.

However, filtering of the electrode signal changes both the SNR-wall
and the SNR which might allow detection of conscious changes. We are
now discussing different filtering approaches. The calculation of both
the SNR and SNR-wall is identical to the methodology described above
with the exception that we assume that subjects can only
\textsl{reduce} consciously an EEG frequency band by a certain
percentage (``desynchronisation'') which is set to 40\% thus the
signal power for the SNR calculation is set to
$T_f(c) = (60\% \cdot 19\mu\mathrm{V})^2$. We are now showing the
impact of filtering on the SNR-wall and SNR:
\begin{itemize}
\item\textbf{$0.1-3~\mathrm{Hz}$}, Fig.~\ref{jaw}D-F: This low frequency bandpass
  filter covers the frequency range of eye-blinks and saccades. While in the original
  trace Fig.~\ref{jaw}A these have not been apparent, now they
  are easily identifiable in Fig.~\ref{jaw}D. In particular during the first 10 seconds the
  subject blinked repeatedly which has been marked with the ``*''. These
  artefacts drive up the difference between chunks of small noise power
  and large noise power which in turn increase the SNR-wall. However, also
  the jaw muscle artefacts show up because of their impulse-like shape
  they trigger large impulse responses of the $0.1-3~\mathrm{Hz}$ bandpass filter which increase
  the maximum variance $\sigma_{r,\textrm{max}}^2$ even further. In contrast the
  SNR stays again negative at $-6~\mathrm{dB}$ while the noise wall is driven up to
  $+14dB$. Consequently, it is not possible to detect conscious changes in EEG.
\item\textbf{$8-18~\mathrm{Hz}$}, Fig.~\ref{jaw}G-I: This moderately
  narrow bandpass filter is a popular choice in BCI applications
  \cite{PFURTSCHELLER2006153} where subjects alter consciously the
  amplitudes of frequencies in this band.  The ratio between lowest
  $\sigma_{r,\textrm{min}}^2$ noise- and highest
  $\sigma_{r,\textrm{max}}^2$ noise-power is comparable to
  Fig.~\ref{jaw}A but the SNR has improved due to less muscle activity
  from the jaw muscles which is only prominent above 20~Hz. Still the
  SNR lies below the SNR-wall and thus detection of conscious EEG
  changes are not possible.
\item\textbf{$8-12~\mathrm{Hz}$}, Fig.~\ref{jaw}J-L: This bandpass filtered signal
  stays substantially below the 20~Hz mark which
  promises less muscle activity. Indeed, the effects of the jaw clenches
  are clearly strongly suppressed in Fig.~\ref{jaw}J.
  This reduces the noise wall to $5dB$ and improves the SNR to $7dB$ which means
  that it is possible to detect conscious alpha frequency changes even with the
  jaw muscles being active.
\item\textbf{$d/dt$}, Fig.~\ref{jaw}M-O: Applying the derivative to the electrode
  signal is a popular choice in BCI as well \cite{duvinage2012biosignals}. The
  derivative or perhaps higher order derivatives act as a highpass filter which
  removes DC and lower frequencies such as eye-movements or eye blinks but
  favours the EMG frequency band. Indeed, when comparing the full range
  electrode signal in Fig.~\ref{jaw}A with the highpass filtered one in Fig.~\ref{jaw}M
  the jaw muscle activity is strongly emphasised. This leads to a high
  SNR-wall of $12dB$ and a low SNR at $-11dB$ and making it impossible to
  detect any conscious EEG changes in the presence of strong non-stationary muscle
  activity.
\end{itemize}

\begin{figure}[!hbt]
\begin{center}
\mbox{\includegraphics[width=\textwidth]{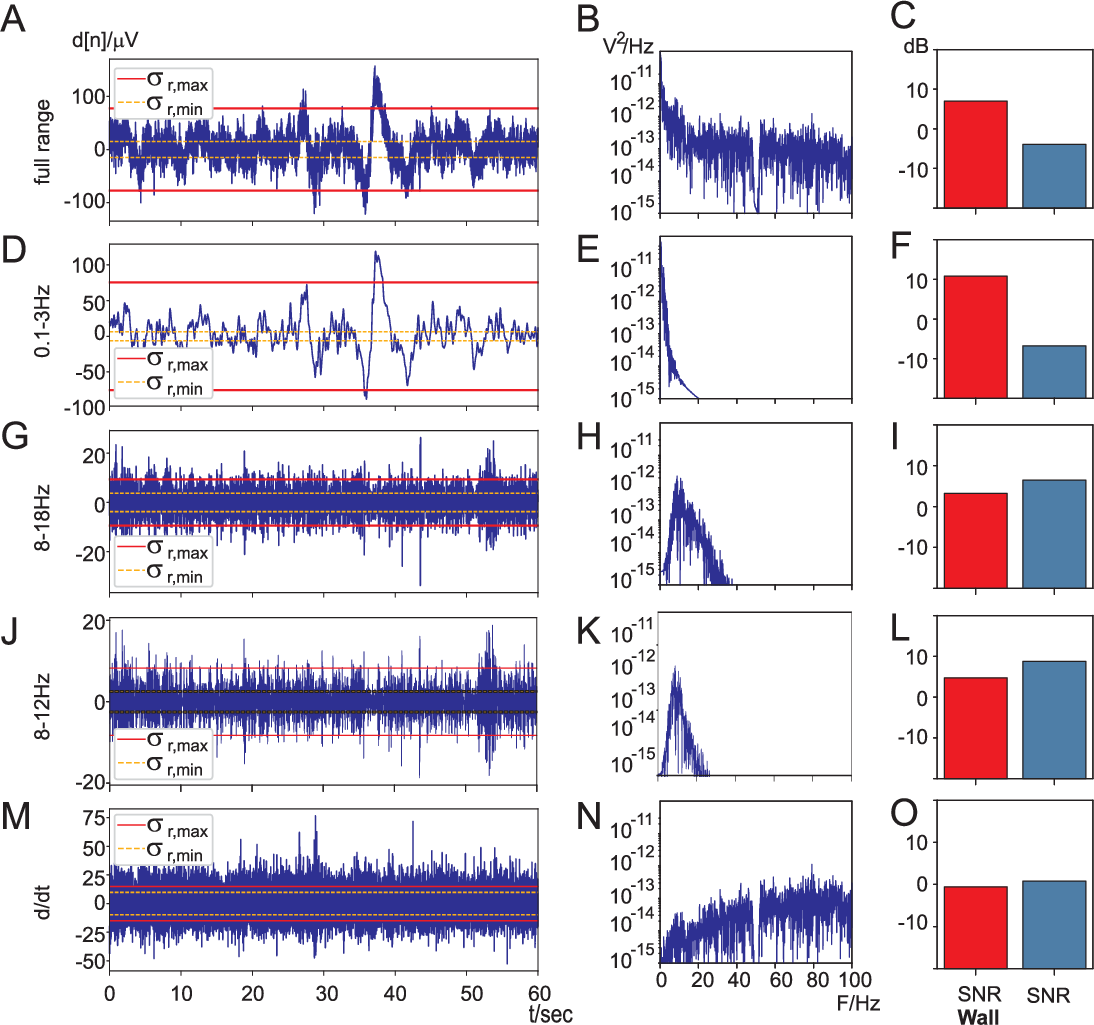}}
\end{center}
\caption{\textbf{SNR and SNR-wall calculations for the ``reading'' task.} Columns A-M:
  electrode signals in the time domain with lines for the maximum variance $\sigma_{r,\textrm{max}}^2$
  and minimum variance $\sigma_{r,\textrm{min}}^2$ while using a sliding window of $2~\textrm{sec}$.
  Columns B-N: frequency power spectra
  of the corresponding time domain signals of columns A-M. Columns C-O:
  SNR-wall and SNR calculations of the corresponding time domain
  signals of columns A-M.
  \label{read}}
\end{figure}
While the jaw clench is excellent for demonstrating the effect of
strong non-stationary muscle noise the reading task is more realistic
in that it contains non-stationary muscle activity at various levels,
eye movements and also movement artefacts. Fig.~\ref{read} follows the
same layout and the same parameters as Fig.~\ref{jaw}. The full range
signal (Fig.~\ref{read}A-C) again exhibits a large difference between
the lowest noise power $\sigma_{r,\textrm{min}}^2$ and the highest
noise power $\sigma_{r,\textrm{max}}^2$ which leads to a high SNR-wall
of $7~\mathrm{dB}$ and at an SNR of $-4~\mathrm{dB}$ it is not
possible to detect conscious EEG changes here. We now move on to the
filtered electrode signals (Fig.~\ref{read}D-O). The filtering of the
electrode signal between $0.1-3~\mathrm{Hz}$ emphasises eye movements
and movement artefacts and shows numerous saccade like voltage changes
which are characteristic of reading where the eyes need to scan the
page. The large spike around $40~\mathrm{sec}$ is a movement
artefact. Together these artefacts cause a large difference between
the smallest noise power and maximum noise power which in turn leads
to a noise wall which is again much higher than the SNR
(Fig.~\ref{read}F). The frequency range of $8-18~\mathrm{Hz}$ is the
one used in motor imagination and rejects most of the muscle
noise. This results in a lower SNR-wall and higher SNR
(Fig.~\ref{read}I) so that now detection of conscious EEG changes are
possible. Similarly, the frequency range of $8-12~\mathrm{Hz}$ for
alpha waves allows detection of conscious EEG changes
(Fig.~\ref{read}L). A curious case is the use of the derivative
(Fig.~\ref{read}M-O) which has been detrimental during the jaw
muscle contractions but here leads to a low SNR-wall and an SNR which
is just above the SNR-wall, potentially allowing the detection of conscious changes of EEG.
This could have been of course just chance
and different for the other subjects which illustrates that both
SNR and SNR-wall are \textsl{random variables} where every subject
in an experiment generates individual SNR-wall \& SNR pairs. This calls now for a
statistical evaluation with all subjects and a t-test which tests for
the SNR being significantly above the SNR-wall which is described
below.

\begin{figure}[!hbt]
\begin{center}
\mbox{\includegraphics[width=\textwidth]{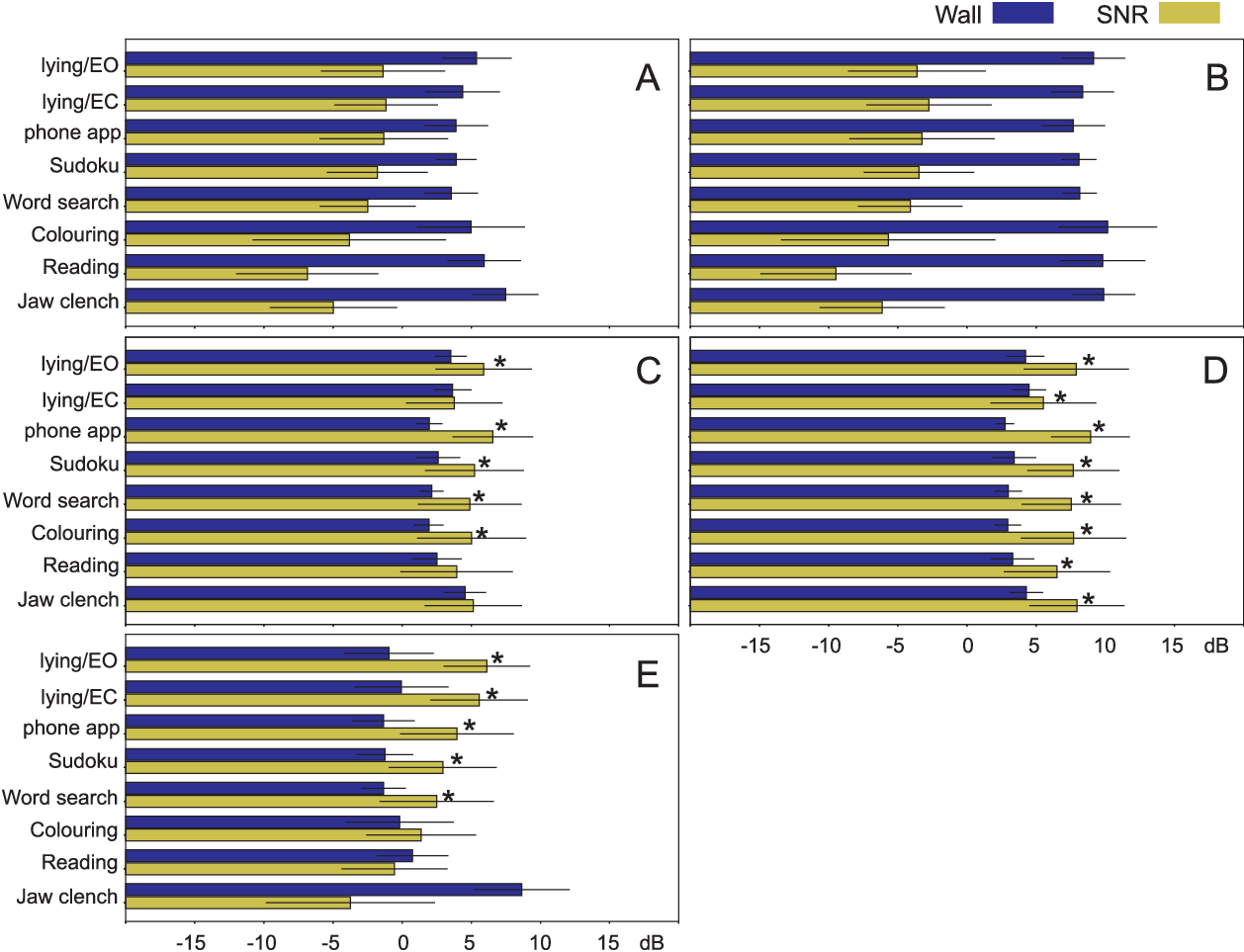}}
\end{center}
\caption{\textbf{Statistical results.} SNR versus SNR-wall for A) full range electrode signal
  with only highpass filtering at $0.1~\mathrm{Hz}$ and 50~Hz removal,
  B) $0.1-3\mathrm{Hz}$ bandpass filtered, C) $8-18~\mathrm{Hz}$ bandpass filtered, D) narrow
  bandpass $10\pm 2~\mathrm{Hz}$ and E) derivative. The bars show the average for
  the SNR-wall and the SNR with error bars for standard deviation. A t-test was used
  to determine if conscious EEG changes can be detected at $p<0.05$ and the ``*'' identifies
  the experimental and filtering conditions where this is significantly possible. \label{results}}
\end{figure}

\subsection*{Statistical analysis}
Fig.~\ref{results} shows the statistical evaluation of the SNR-walls
against the SNRs. Each panel Fig.~\ref{results}A-E shows the SNR-walls
and SNRs for the different tasks ranging from lying down to jaw
clenching. Panels A-E themselves represent different filtering
methods of the electrodes signals. Both SNR and SNR-walls are shown in
dB on a scale from -15~dB to +15~dB. The y-axis shows the results for
each separate task so that one can decide if it is possible to detect
a conscious change in the EEG reliably or not. Both SNR and SNR-walls
are random variables from the different participants and a t-test
($p<0.05$) was used to determine if the SNR is significantly above the
SNR-wall which then predicts that conscious EEG changes can be
detected for a specific experiment using a specific causal filtering
technique. We describe the statistical analysis first against
the different pre-processing methods and then in relation to the
different tasks.

\begin{description}
  \item[\textbf{A) $0.1>~\mathrm{Hz}$:}]
Fig.~\ref{results}A shows the SNR/SNR-wall
pairs after only highpass filtering above $0.1~\mathrm{Hz}$ and mains
removal. The task ``jaw clench'' has an average SNR-wall of
$8~\mathrm{dB}$ but given that the SNR is at $-5~\mathrm{dB}$ it is
significantly not possible to detect conscious EEG here. Generally in
Fig.~\ref{results}A all average SNR values are lower than the average
SNR-wall values and the corresponding t-test reports that electrode
signals after minimal filtering significantly cannot detect a
conscious change of EEG at all.
\item[\textbf{B) $0.1-3~\mathrm{Hz}$:}]
As discussed in the
previous sections this frequency range strongly emphasises eye
movements and movement artefacts which leads to high average SNR-walls
and very low SNR values for all tasks. Consequently the t-test
provides again no significant case where conscious EEG changes could
be detected.
\item[\textbf{C) $8-18~\mathrm{Hz}$:}]
This bandpass filter
yields significant results predicting the detection of conscious EEG
changes for: lying (eyes closed), using a phone app, playing Sudoku,
doing word search and colouring. Reading which engages a large amount
of facial muscles and the jaw clench as a worst-case scenario
obliterate any detection effort. It is interesting to see that lying
down with eyes closed won't allow detection which anecdotally points
to the subjects being more tense having a higher
muscle activity.
\item[\textbf{D) $8-12~\mathrm{Hz}$:}]
Best detection ability is achieved with a narrow
frequency bandpass around the alpha band with
all tasks allowing detection of a consciously changed EEG.
\item[\textbf{E) $d/dt$:}]
Here, 
pre-filtering is done by a derivative which is popular for
realtime BCI \cite{duvinage2012biosignals}. Here, lying down with eyes
closed and open, using a phone app, playing Sudoku, doing word search
and colouring allow the detection of a conscious EEG
change. Colouring, reading and jaw clench have the highest EMG
components and fail the t-test which means that conscious EEG changes
cannot be detected. As already observed earlier the derivative
emphasises EMG activity which is apparent in panel E for jaw clench
where the average SNR is about 10~dB lower than the noise wall and
creates a clear cut non-significant result.
\end{description}
It is interesting to note
that with the derivative keeping in particular EMG intact that there
is a clear hierarchy of SNR values from eyes closed having the highest
SNR down to the jaw clench for the lowest. In contrast, the bandpass filter
approaches in panels C) and D) do not have this dependence with both
bandpass ranges below the 20~Hz boundary for EMG confirming again that
keeping BCI detection below 20~Hz is the safest bet to avoid EMG
interference.

\begin{table}[!ht]
\centering
\caption{
  {\bf Detection of conscious EEG changes significantly possible ordered by task.}}
\begin{tabular}{l|l|l|l|l|l|}
\hline
            & \textbf{A} $0.1>~\mathrm{Hz}$ & \textbf{B} $0.1-3~\mathrm{Hz}$ & \textbf{C} $8-18~\mathrm{Hz}$ & \textbf{D} $10\pm 2~\mathrm{Hz}$ & \textbf{E} $d/dt$\\ \hline
lying/EO    &   &   & * & * & * \\ \hline
lying/EC    &   &   &   & * & * \\ \hline
phone app   &   &   & * & * & * \\ \hline
Sudoku      &   &   & * & * & * \\ \hline
Word Search &   &   & * & * & * \\ \hline
Colouring   &   &   & * & * &   \\ \hline
Reading     &   &   &   & * &   \\ \hline
Jaw clench  &   &   &   & * &   \\ \hline
\end{tabular}
\label{statstasks}
\end{table}
In the previous section we have ordered the results according to the \textsl{pre-filtering} of the electrode signal.
Of equal interest is which \textsl{task} allows the detection of conscious EEG changes and which
renders it impossible to do so. For this purpose we have re-ordered the results
of Fig.~\ref{results} and show them task by task in
table~\ref{statstasks}. As in Fig.~\ref{results} the ``*'' in
table~\ref{statstasks} indicate that conscious EEG changes can be
significantly detected. The labels A-E match the panel labelling in
Fig.~\ref{results} and are the different pre-filtering methods.
\begin{description}
\item[\textbf{Lying down, eyes open:}]
  Conscious detection of EEG is possible here for moderate wideband prefiltering ($8-18~\mathrm{Hz}$), narrow alpha band
  filtering ($10\pm 2~\mathrm{Hz}$) or using the derivative which acts as a highpass.
  Lying flat down and having the eyes open mainly creates
  eye blink artefacts which, without any filtering, cause a high ratio between the minimum and maximum noise variances
  which in turn result in a high SNR wall. EEG detection is possible if these eye-blink artefacts are removed by a highpass filter
  characteristic.
\item[\textbf{Lying down, eyes closed:}]
  With lying down and eyes closed detection of conscious EEG changes is still possible when focusing on the narrow alpha band
  or using the derivative as a highpass filter. However wider bandpass filtering yields no significant results. As mentioned above
  closing the eyes in a lab is slightly uncomfortable which seems to result in a higher muscle tone. Comparing
  Fig.~\ref{results}C between eyes open and closed only the SNR changes but not the SNR-wall which means that the actual
  ratio of min and max noise variance is not different between those two tasks but rather the constant muscle tone which
  is reflected in a higher SNR while having the eyes closed.
\item[\textbf{Phone app, Sudoku, Word search:}] As outlined further above these three tasks have only moderately elevated non-stationary
  muscle activity and eye-movements. The muscle activity mostly arises from being a bit more tense when solving the tasks
  but causes only low ratios between highest noise variance and lowest noise variance which in turn results in low SNR-walls.
  Even the simple derivative which favours muscle noise still yields significant results. Again, the low frequency
  filtering and wideband filtering make it impossible to detect conscious EEG changes because the low frequency eye
  movements cause high SNR-walls.
\item[\textbf{Colouring:}] This is now a task which involves strong arm muscle activity but also most likely compensatory
  muscle activity in the entire body such as neck muscles.
  This results in higher muscle noise levels compared to the previous tasks and now the filtering
  with the derivative no longer allows detection of conscious EEG changes. However, wideband prefiltering ($8-18\mathrm{Hz}$) and
  narrow alpha band filtering ($10\pm 2~\mathrm{Hz}$) still allow significant EEG detections.
\item[\textbf{Reading aloud:}] This task causes irregular contraction of facial muscles and the eyes need to scan
  the text which cause eye artefacts as well. This non-stationary noise leads to high SNR walls which
  prevent the detection of conscious EEG changes except when performing narrow band filtering around the alpha band. This
  filtering steers clear of both eyeblink artefacts and higher frequency muscle noise as much as possible.
\item[\textbf{Jaw clenching:}] The jaw muscles create very high amplitude bursts of muscle activity which are about
  50 times higher than the baseline noise. Without any filtering this results in high ratios of minimum and maximum
  noise variances which in turn results in very high noise walls. Similarly to reading aloud removing the non-stationary
  muscle noise is crucial to be able to detect conscious EEG changes and is
  only possible if detection focuses just on the narrow alpha band.
\end{description}
Overall tasks with strong non-stationary facial muscle activity such
as reading aloud or jaw clenching allow only significant detection of
conscious EEG changes if one performs narrow band filtering around the
alpha band which steers clear as much as possible of both muscle noise
and eye movement artefacts. However, eye movement artefacts are less
of a problem than muscle artefacts because their frequencies do not
overlap with the standard BCI frequency detection ranges. Those tasks
which contain mostly eye movements but little muscle noise such as
playing a video game allow detection of conscious EEG changes with a
wideband BCI bandpass ($8-18\mathrm{Hz}$) but also with a derivative
which only attenuates eye-movements but not muscle noise.

\section*{Discussion}
In this paper we have introduced an objective hard criterion which
 determines if it is possible to detect conscious EEG changes in a
 recording contaminated with non-stationary noise. Specifically, this
 methodology requires the SNR of an electrode signal to be higher than
 its SNR-wall so that conscious changes in EEG can be detected. The
 SNR-wall is calculated by taking the ratio between the minimum
 and maximum noise variances of the electrode signal
 (Eq.~\ref{snrwall}). The SNR is calculated using the standard
 definition of SNR which is here the ratio between the consciously generated EEG
 component against the nominal noise level (Eq.~\ref{snrdef}). Both
 equations result from sound analytical calculations and were then used to
 calculate all SNR-wall and SNR values for all experimental conditions and tasks.
 In other words, no manual selections of the SNR-walls, SNRs or thresholds are
 needed for the BCI-walls framework protecting it from subjective, possibly arbitrary
 choices. We then determined the SNR-wall for a range of different
tasks with non-stationary noise while applying different filtering
techniques. Wideband filtering and filtering in the delta band make it
impossible to detect any conscious EEG changes. On the other hand
filtering around the alpha band allows EEG detection even under high
non-stationary muscle interference during jaw clenching.

As outlined in the introduction, EEG is contaminated with different
forms of noise where EMG is the hardest to remove because of its broad
frequency spectrum \cite{Goncharova2003} ranging from 20 to 80~Hz. The
gold standard to measure the EMG contribution is by neuromuscular
blockade where a subject is temporally paralysed and, thus, generates
zero EMG. Measuring brain activity from a paralysed subject reveals
substantial EMG contamination above 20~Hz \cite{Whitham2007} and
matches our results where the SNR-wall is lowest if one uses a
narrow bandpass around the 10~Hz alpha band \cite{Wolpaw1991} steering
clear of the EMG spectrum.

An existing method of how to determine if EMG interference is
detrimental to EEG detection is to use the amplitude of a bandpass
filtered electrode signal and suspend data collection when it is above
a certain threshold \cite{GONCHAROVA20031580}. Another suggested
method is to use a montage consisting of a large number of electrodes
and then apply independent component analysis (ICA) to the electrode
signals
\cite{GONCHAROVA20031580,jung_makeig_humphries_lee_mckeown_iragui_sejnowski_2000,Delorme2007}
which separates noise from EEG which then allows the calculation of
the SNR. The ICA could also be used to determine the SNR for the
BCI-wall methodology. Instead, we have chosen the P300 methodology to
determine the pure consciously controlled EEG power because ICA
requires by definition more than one electrode and we recorded from
just one. Future research could also calibrate the SNRs obtained from
the ICA with those from the P300 responses. Be it P300 or ICA they are in
stark contrast to determining the SNR by paralysing a participant
temporarily \cite{Whitham2007,Fitzgibbon2013} which would certainly
yield the best results but no doubt would raise substantial ethical
concerns.

So far ICA has only been suggested to determine the SNR but it can
also be used to determine the SNR-wall because the minimum
$\sigma_{r,\textrm{min}}^2$ and maximum noise variances
$\sigma_{r,\textrm{max}}^2$ are readily available in the independent
component(s) of the ICA carrying the noise. Since the ICA is a 
well-established standard tool for EEG offline analysis calculating both
the SNR and SNR-wall could just be a simple additional step which then
allows the robust evaluation of an experiment.

However, it appears that most BCI studies and reviews
\cite{Zhang2021,Kundu2021} stay silent about how they have dealt with
noise interference. The first comprehensive literature survey
investigating artefact removal \cite{FATOURECHI2007480} finds that
most BCI papers do not report whether or not they have considered the
presence of EMG (67.6\%) or EOG (53.7\%) artefacts in brain
signals. This situation has not improved much ten years later
\cite{Craik_2019} where 41\% of the studies did not mention any
artefact removal process of EEG and even where artefacts were
mentioned 22\% of those studies did not do any cleaning or artefact
removal. Given that the BCI-walls methodology has been proven to be
analytically sound \cite{Tandra2008} and requires only standard EEG
recording parameters such as signal- and noise-variances it should be
part of a standard process to evaluate whether reliable EEG detection is
possible or not. Pre-registrations of experimental designs could also
specify a BCI-wall test which could then later provide a robust decision
point as to whether EEG recordings have been sufficiently high quality to allow the
detection of conscious EEG changes or if the data needs to be
discarded.

SNR-wall theory was developed in the field of telecommunications \cite{Tandra2008} where
it is common that a multitude of transmitters create non-stationary
background noise
\cite{Tani2022,9706147,Gohain2018,Vivekanand2021,Captain2019,Qian2015}
and, for example, making it impossible for a mobile phone to communicate
with its base station with an increasing number of transmitters.
The SNR-wall calculations are able to make strong predictions
about the capability of a telecommunications system to work as intended,
and similarly this should be also standard practise for BCI systems.

\section*{Funding Statement}
This work was supported by the School of Engineering, University of Glasgow.

\section*{Conflict of Interest Statement}
B.P. is CEO of Glasgow Neuro LTD which manufactures the Attys DAQ board. 
This does not alter our adherence to PLOS ONE policies on sharing data 
and materials.

\section*{Acknowledgements}
We thank Nicholas J Bailey for his constructive feedback on the manuscript.

%\nolinenumbers

%\bibliography{bci,snrbci}

\end{document}